\documentclass[journal]{vgtc}                

\ifpdf
  \pdfoutput=1\relax                   
  \usepackage{graphicx}                
  \DeclareGraphicsExtensions{.pdf,.png,.jpg,.jpeg} 
\else
  \usepackage{graphicx}                
  \DeclareGraphicsExtensions{.eps}     
\fi%

\graphicspath{{figures/}{pictures/}{images/}{./}} 

\usepackage{microtype}                 
\PassOptionsToPackage{warn}{textcomp}  
\usepackage{textcomp}                  
\usepackage{mathptmx}                  
\usepackage{times}                     
\usepackage{cite}                      
\usepackage{tabu}                      
\usepackage{booktabs}                  
\usepackage{multirow}

\usepackage{amssymb}    
\usepackage{paralist}  

\usepackage{tabularx}  
\usepackage{wrapfig}  

\usepackage{threeparttable}  
\usepackage{xcolor}
\usepackage{nth}

\usepackage{listings}
\definecolor{mygray}{rgb}{0.8,0.8,0.8}
\usepackage{setspace} 
\usepackage{hyperref}

\onlineid{1346}

\vgtccategory{Data Transformations}
\vgtcpapertype{Data Transformations}

\newcommand{\somnus}{{\scshape Somnus}}
\newcommand{\name}{{\scshape Comantics}}

\title{Revealing the Semantics of Data Wrangling Scripts \\ With \name{}}

\author{Kai Xiong, Zhongsu Luo, Siwei Fu, Yongheng Wang, Mingliang Xu, Yingcai Wu}
\authorfooter{
\item 
  K. Xiong and Y. Wu are with the State
  Key Lab of CAD\&CG, Zhejiang University, Hangzhou, China, and with the Zhejiang Lab, Hangzhou, China.
  E-mail: \{kaixiong, ycwu\}@zju.edu.cn.
\item
Z. Luo is with Zhejiang University of Technology, Hangzhou, China, and with the Zhejiang Lab, Hangzhou, China. E-mail: rickyluozs@gmail.com.
\item
S. Fu and Y. Wang are with the Zhejiang Lab, Hangzhou, China. E-mail: fusiwei339@gmail.com, wangyh@zhejianglab.com.
\item 
M. Xu is with the School of Computer and Artificial Intelligence, Zhengzhou University, Zhengzhou, China.
 E-mail: iexumingliang@zzu.edu.cn.
\item
Yongheng Wang and Siwei Fu are the co-corresponding authors.
}

\shortauthortitle{Biv \MakeLowercase{\textit{et al.}}: Global Illumination for Fun and Profit}

\abstract{
  Data workers usually seek to understand the semantics of data wrangling scripts in various scenarios, such as code debugging, reusing, and maintaining. However, the understanding is challenging for novice data workers due to the variety of programming languages, functions, and parameters. Based on the observation that differences between input and output tables highly relate to the type of data transformation, we outline a design space including $103$ characteristics to describe table differences. Then, we develop \name{}, a three-step pipeline that automatically detects the semantics of data transformation scripts. The first step focuses on the detection of table differences for each line of wrangling code. 
  Second, we incorporate a characteristic-based \kai{component} and a Siamese convolutional neural network-based component for the detection of transformation types.
  Third, we derive the parameters of each data transformation by employing a ``slot filling'' strategy. 
  We design experiments to evaluate the performance of \name{}. Further, we assess its \kai{flexibility} using three example applications in different domains.
} 


\CCScatlist{ 
 \CCScat{K.6.1}{Management of Computing and Information Systems}%
{Project and People Management}{Life Cycle};
 \CCScat{K.7.m}{The Computing Profession}{Miscellaneous}{Ethics}
}

\teaser{
  \centering
  \includegraphics[width=\linewidth]{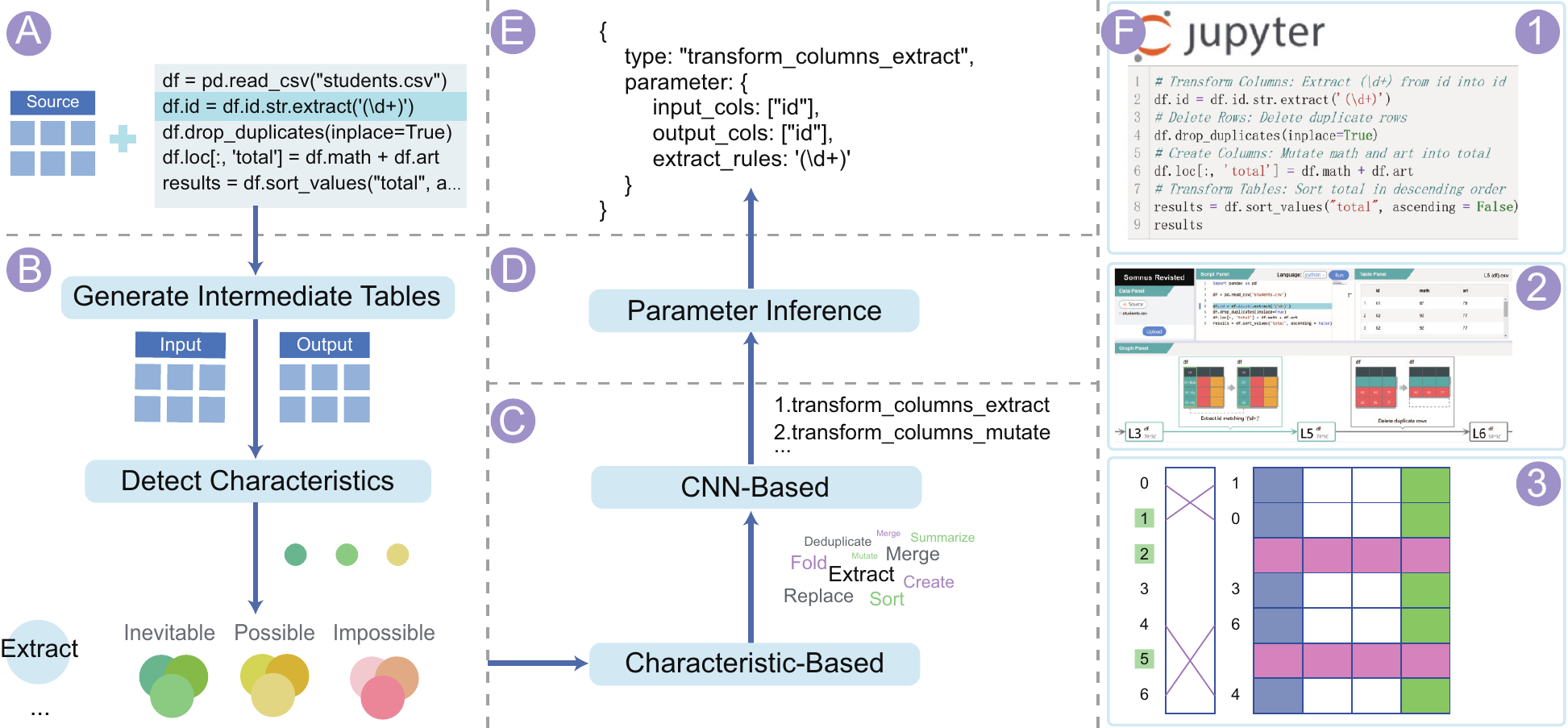}
  \caption{\name{} is a pipeline that reveals the semantics of wrangling code. (a) shows the input of the pipeline, \ie, a data table and a piece of wrangling script. (b) is the first step of \name{}, aiming to generate intermediate tables for each code and detect changes between tables. (c) depicts the second step that identifies transformation type with characteristic-based and CNN-based components. (d) is the third step that infers parameters for the transformation. (e) shows the output of \name{}. (f) shows that our pipeline is applied to three applications in different domains.}
	\label{fig:teaser}
}

\vgtcinsertpkg

\newcommand{\etal}{et~al.\ }
\newcommand{\eg}{e.g.}
\newcommand{\ie}{i.e.}

\definecolor{kLime}{HTML}{02b340}
\definecolor{OrangeRed}{HTML}{FF4500}
\newcommand{\kai}[1]{\textcolor{black}{#1}}

\definecolor{fswGreen}{HTML}{70AD47}
\newcommand{\green}[1]{\textbf{\textcolor{fswGreen}{#1}}}

\definecolor{fswOrange}{HTML}{ED7D31}
\newcommand{\orange}[1]{\textbf{\textcolor{fswOrange}{#1}}}

\begin{document}


\firstsection{Introduction}

\maketitle



Data wrangling is an arduous and time-consuming routine~\cite{dasu2003exploratory, guo2011proactive} for data workers who carry out data analysis tasks with different levels of expertise~\cite{liu2019understanding, bartram2022untidy}. 
A commonplace method of wrangling data is writing custom cleaning scripts in various programming languages~\cite{kandel2011wrangler, krishnan2017boostclean}, such as R and Python.
In this case, data workers usually seek to understand the semantics of wrangling scripts in various scenarios~\cite{yang2021subtle}, including debugging for potential code issues, reusing scripts for their data, and maintaining code that is not well-documented.
Here, the \textit{semantics of wrangling scripts} means the \green{type of data transformation} and its \orange{parameters}.
For example, given a line of code written in R, \lstinline|tb2 = arrange(tb1, -num)|, the semantics of the code refers to \green{sorting} the column \orange{num} of the input table \orange{tb1} in \orange{descending} order, and returning the output table \orange{tb2}.

However, understanding the semantics of intricate scripts requires advanced programming skills and is error-prone~\cite{lewis1987can, sorva2013review, qian2017students}. 
Novice data workers may find it challenging for three reasons.
First, data workers need to work across programming languages in some scenarios, such as learning a new programming language and reusing code written in another language. 
However, different scripting languages have diverse regulations in implementing data transformations.
For example, the \textit{sort} transformation can be implemented by the function \lstinline|arrange| in \kai{dplyr~\cite{wickham2021dplyr} (R)}, and the descending order is represented by a minus sign or the \lstinline|desc| function. 
While in \kai{Pandas~\cite{reback2022pandas} (Python)}, the same transformation is implemented by the function \lstinline|sort_values|, and the descending order is noted as \lstinline|ascending=False|.
Second, in one language, there are different functions to implement the same data transformation.
For instance, both \lstinline|gather| and \lstinline|pivot_longer| in \kai{tidyr~\cite{wickham2020tidyr} (R)} can perform the \textit{fold} transformation, and there is a subtle difference in their usage.
Third, some functions  may omit important parameters for simplicity.
These functions either have default parameters (\eg, the \lstinline|sort_values| function in \kai{Pandas} performs sorting in ascending order by default), or can extract parameters from the input data tables 
(\eg, the \lstinline|left_join| function in \kai{dplyr} has built-in rules to infer join keys from the two input tables).

A body of prior work has been proposed to understand wrangling scripts.
Some tools such as Unravel~\cite{shrestha2021unravel} and WrangleDoc~\cite{yang2021subtle} are used for exploring and debugging wrangling scripts.
These tools are useful for revealing changes in data tables after performing a data transformation.
However, they cannot directly inform data workers of the type of transformation.
Others, such as \somnus{}~\cite{kai2022somnus}, Datamations~\cite{pu2021datamations}, and Data Tweening~\cite{khan2017data}, illustrate the semantics of data transformations using well-designed visualization.
However, they rely on hand-crafted rules to parse scripts, resulting in poor generalizability and scalability.


In this paper, we propose a pipeline, called \name{} meaning \textbf{co}de se\textbf{mantics}, that identifies the semantics of wrangling scripts. 
We observe that, given a line of wrangling code, changes between input and output tables highly relate to the type of transformation. 
To have an overview of the changes, we summarize the characteristics of table changes by analyzing a corpus containing $921$ lines of real-world wrangling code and present a design space comprising two primary dimensions, i.e., the types of data objects and the properties of data changes. 
Then, we construct the pipeline consisting of three key steps, \ie, Data Preprocessing, Type Inference, and Parameter Inference.
For Data Preprocessing, \name{} executes the script and obtains intermediate data tables for each line of wrangling code.
Then, the characteristics of table changes are identified automatically.
The Type Inference step consists of two components. 
The first component infers transformation type based on the characteristics and obtains a list of candidate transformation types.
To identify the transformation with maximum possibility, we employ a Siamese convolutional neural network\cite{Aditya2019Siamese} to rank transformation candidates.
Based on the inferred transformation, Parameter Inference aims to identify parameters for transformations. 
We borrow the ``slot filling'' strategy and detect parameters based on the wrangling script, table contents, and characteristics.
Finally, \name{} outputs a data transformation with maximum likelihood and its parameters, which can be applied to a variety of downstream tasks.

We design experiments to evaluate the performance of \name{}. 
We first annotate the real-world corpus with data transformations. Then we assess the generalizability across different programming languages and the contribution of different components of \name{} based on five experiment settings.
To evaluate the \kai{flexibility} of our pipeline, we apply it to three applications in different domains, \ie, improving Jupyter Notebook with automatic annotation, augmenting \somnus{}~\cite{kai2022somnus} with scalable backend, and enhancing TACO~\cite{niederer2017taco} with additional information.

In summary, the contributions of this paper are four-fold:
First, we summarize a design space including $103$ characteristics of table changes.
Second, we construct a novel pipeline, \name{}, for inferring semantics, \ie, the types of transformations and their parameters, of wrangling scripts.
Third, we evaluate the performance of our pipeline using quantitative experiments.
A sub-contribution is that we have built a real-world dataset including $921$ lines of wrangling code where each is annotated with a data transformation.
Fourth, to assess the \kai{flexibility} of \name{}, we apply our pipeline to three example applications in different domains.

\section{Related Work}

In this section, we discuss techniques from three aspects, \ie, data wrangling, program understanding in data wrangling, and table comparison.

\subsection{Data Wrangling}

Data wrangling is a nontrivial activity of cleaning, transforming, and enriching data into the desired form palatable for downstream tasks~\cite{kandel2011wrangler, guo2011proactive}.
Data wrangling often involves a significant number of data transformations~\cite{rattenbury2017principles}.
Kasica \etal\cite{kasica2020table} developed a concise and actionable framework describing multi-table transformation operations.

There are two kinds of approaches for facilitating the process of data wrangling.
On the one hand, a plethora of interactive applications has been proposed.
Tools for wrangling tabular data~\cite{HiTailor2022Li, Rigel2022chen} like Microsoft Excel, Tableau Prep Builder~\cite{Tableau2019}, OpenRefine~\cite{OpenRefine}, and Trifacta~\cite{Trifacta} provide an interactive menu for selecting the desired transformation.
Further, Trifacta and its predecessor Data Wrangler \cite{kandel2011wrangler,guo2011proactive} integrate an inference engine  that generates a ranked list of suggested transformations.
Others aim to address wrangling graph~\cite{liu2011network, heer2014orion, bigelow2019origraph, tominski2021toward, hu2022visualizing, mei2020datav} and website~\cite{lin2009end, morcos2015dataxformer,abedjan2016dataxformer, inala2017webrelate, liu2021muluba} data.
Some of the above tools~\cite{Tableau2019, OpenRefine, Trifacta, kandel2011wrangler, guo2011proactive} support recording the process of data transformations through a domain-specific language, and provide textual descriptions of transformations.
On the other hand, numerous scripting languages (\eg, Python and R) with data manipulation libraries (\eg, Pandas~\cite{reback2022pandas}, dplyr~\cite{wickham2021dplyr}, tidyr~\cite{wickham2020tidyr}, and tidyverse~\cite{Wickham2019tidyverse}) have been commonly used to wrangle data for their flexibility.
However, programming is a complicated and burdensome activity, requiring developers to master specialized skills~\cite{lewis1987can, pane2002programming}.
To reduce this barrier, increasing research work~\cite{jin2017foofah, drosos2020wrex, feng2017component, he2018transform} is motivated by \textit{programming by example} that synthesizes target code given examples provided by data workers.
Compared to the aforementioned approaches, our pipeline has different goals, which focus on helping data workers understand wrangling scripts.

\subsection{Program Understanding in Data Wrangling}
Program understanding in data wrangling is a hot research topic that aims to help data workers comprehend code.
Prior work roughly falls into two categories.
Firstly, some tools are designed for debugging.
For example, WrangleDoc~\cite{yang2021subtle} is a JupyterLab extension that leverages program synthesis techniques to automatically generate summaries of code fragments to find subtle bugs.
Unravel~\cite{shrestha2021unravel} provides summaries for data transformation functions and an always-on visualization for fluent code to explore and debug the chain of expressions in R.
Besides, TweakIt~\cite{lau2021tweakit} is a system implemented as an extension to Microsoft Excel that applies live interactions to help data workers explore and understand the effects of unfamiliar code.
\name{} shares the same design goal of facilitating the comprehension of wrangling scripts.
However, we focus on the semantics of data wrangling, which consists of the types of data transformations and their parameters. 

Secondly, some tools are proposed to visualize the semantics of data wrangling.
For instance, Datamations~\cite{pu2021datamations} creates animated transitions for data transformations to explain the entire data analysis pipeline.
\somnus{}~\cite{kai2022somnus} supports transformation inference based on function information and generates a glyph-based provenance graph to visualize the evolution of tables. 
These visualization tools are useful and effective in depicting the semantics of data processing workflow.
However, they rely on a rule-based engine to parse the semantics of the wrangling process, which can hardly scale to different programming languages and a variety of functions.
In addition, crafting rules is a laborious process and is limited in real-world applications.
\name{}, on the other hand, infers the semantics of wrangling scripts from input/output tables and individual lines of code. It is independent of programming languages, libraries, and functions, resulting in better scalability and generalizability. 
We have successfully applied \name{} to three applications (see Sec.~\ref{sec:app}) to assess its \kai{flexibility}.

\subsection{Table Comparison}
Data comparison techniques are widely used to investigate differences between multiple versions of data.
Although plenty of work aims to compare time series data, image data, graphs, etc., there are few techniques designed to compare tabular data~\cite{niederer2017taco}.
Available table comparison tools, such as ExcelCompare~\cite{ExcelCompare}, AQT~\cite{AQT}, DiffKit~\cite{DiffKit}, Daff~\cite{daff}, Ridom SeqSphere+~\cite{SeqSphere}, and Compare~\cite{Compare}, identify and compute differences between two input tables according to metrics including table size differences, cell content differences, the number of unique records and fields, etc., and output statistical results, textual summaries, or visualizations. 
Sutton \etal\cite{sutton2018data} presented a tool to detect and explain the differences between two tables and formulate executable patches that can transform one table to be compatible with the other.
In addition, Niederer \etal\cite{niederer2017taco} developed TACO, a visual comparison tool for tabular data. It encodes table differences through interactive visualizations to illustrate different types of changes. 
The aforementioned approaches are successful in comparing differences between tables. 
\name{} goes one step forward in which table differences are identified to infer the semantics of wrangling scripts.

\section{A Design Space for Table Changes}
\label{sec:space}

The semantics of wrangling scripts consists of transformation types and their parameters. 
According to Xiong \etal\cite{kai2022somnus}, identifying the transformation type is the key challenge.
We observe that differences between input and output tables of a line of wrangling code can reveal the transformation type to some extent. 
For example, if some rows are deleted while the others remain unchanged, the transformation seems to belong to \textit{delete rows}.
Further, if all deleted rows contain missing values while the remaining do not, the transformation type is likely to be \textit{delete rows with missing values}.
Motivated by this observation, we explore the characteristics of table changes and present a design space to understand how a table is changed based on data transformations.

\subsection{Methodology}
\label{subsec:space_method}
\begin{figure}[!hbt]
    \centering
    \includegraphics[width=0.9\linewidth]{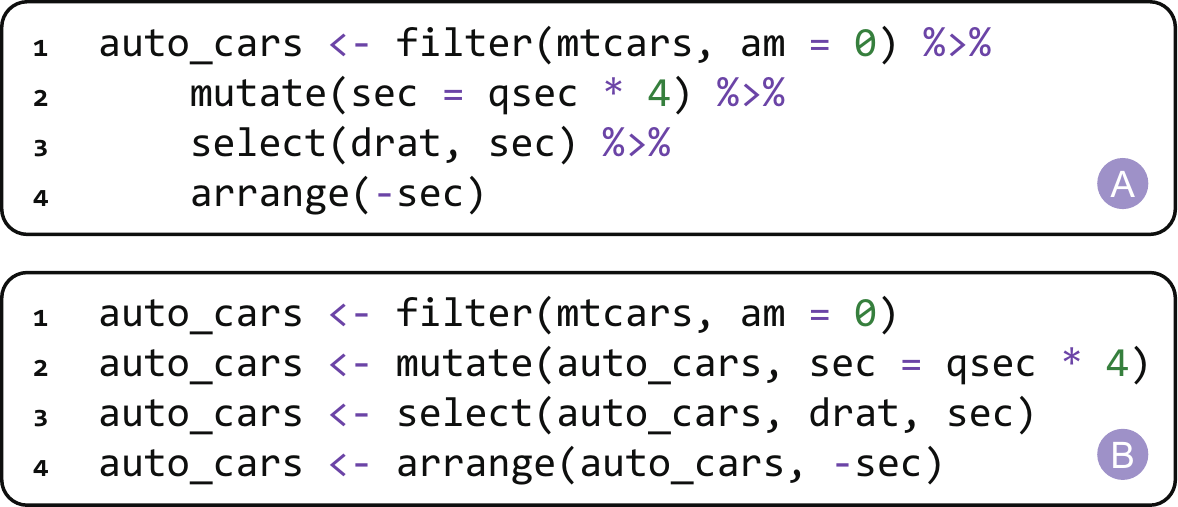}
    \caption{An example of unraveling (a) a chained syntax into (b) a sequence of statements in the R programming language.}
    \label{fig:unravel_chain}
\end{figure}

To understand the characteristics of table changes, we collect real-world corpus and analyze how wrangling scripts transform input data tables into outputs.
The corpus is collected from GitHub\footnote{\url{https://github.com}} and Kaggle\footnote{\url{https://www.kaggle.com}}. 
We use keywords such as data wrangling and data cleaning to find target scripts.
Furthermore, we collect corpus from prior work~\cite{kasica2020table,feng2017component}. 
\kai{The collection procedure is based on two inclusion criteria. First, as our approach relies on table changes, we focus on scripts with data tables. 
Second, due to our expertise in programming, we keep scripts written in Python and R, and balance the number of scripts in both languages.}
\kai{After collecting the corpus, we manually ``clean'' these scripts to meet the requirement of \name{}.
Besides data wrangling, the scripts usually include code for other purposes, such as visualization and machine learning. 
We carefully remove these code and ensure that the remaining can run and process the input data correctly.
Then, we deal with function chaining, where multiple functions are executed consecutively on the same variable.}
An example is shown in Fig.~\ref{fig:unravel_chain}(a), four functions (\ie, \lstinline|filter|, \lstinline|mutate|, \lstinline|select|, and \lstinline|arrange|) are called in sequence on the dataset \lstinline|mtcars|.
This language pattern is efficient for developers. However, it is not supported in \name{} because we can hardly obtain intermediate tables.
Hence, we rewrite chained functions into individual lines of transformations (Fig.~\ref{fig:unravel_chain}(b)).
Finally, our corpus contains $74$ curated scripts, where $42$ of them are written in Python and $32$ are in R, adding up to $921$ lines of wrangling code.

To code the characteristics of table changes, we first execute these scripts to obtain the intermediate tables for each line of wrangling code.
Then we record the differences between tables in detail from various aspects as much as possible.
Next, we apply a qualitative method, open coding and axial coding \kai{rooted in grounded theory}, to these table differences.
Specifically, we label each difference with a descriptive code and draw connections between these codes.
Based on our observations and wrangling experiences, we adopt the codes which are conducive to distinguishing between different transformations.
After frequent discussions with two data scientists, we group and condense related codes into broader categories.
Finally, we derive five properties of data changes and four data objects from these categories.

\subsection{Dimensions of the Design Space}
\begin{figure*}[!t]
	\centering
    \vspace{-1mm}
	\includegraphics[width=0.981\linewidth]{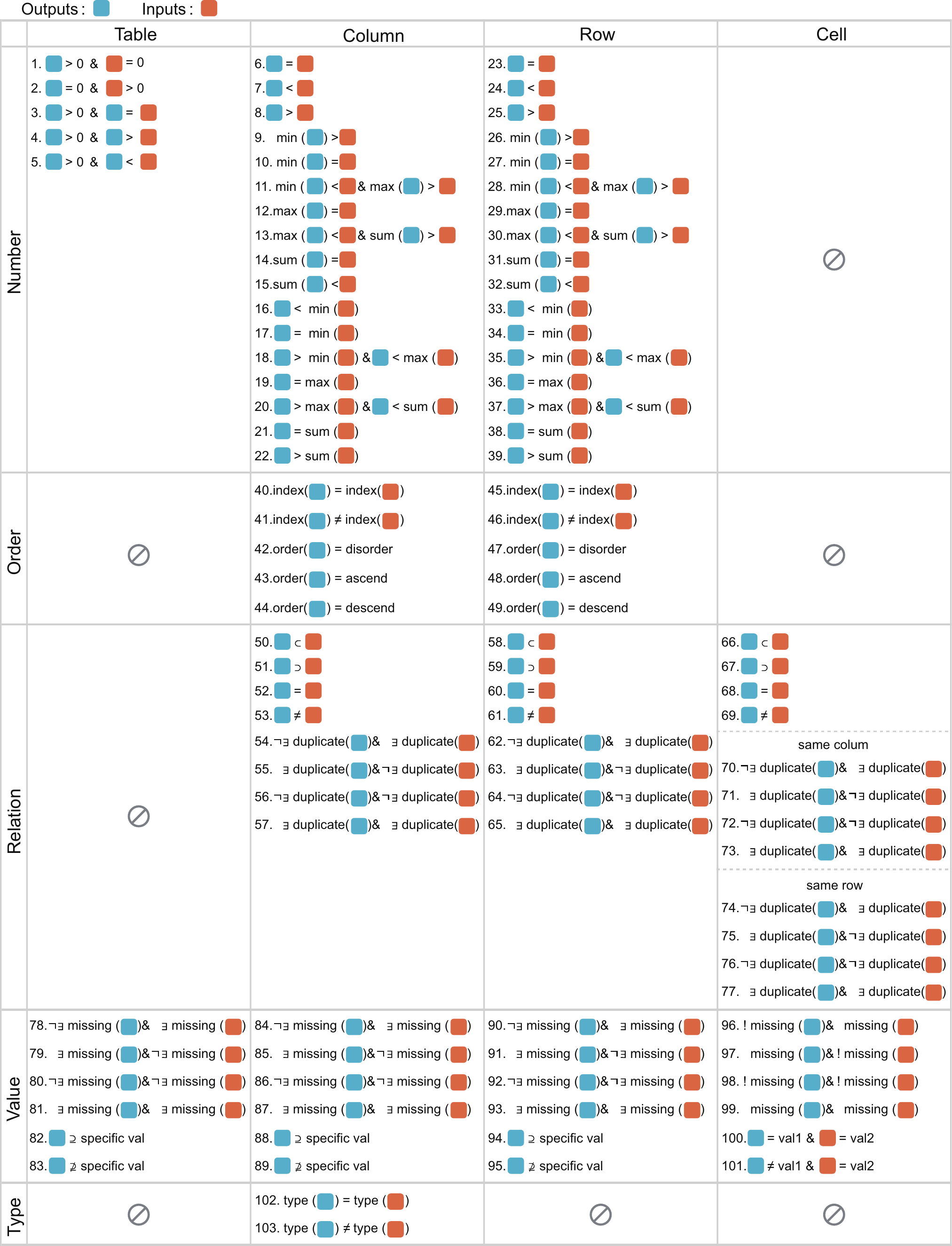}
    \vspace{-1.1mm}
	\caption{
        The design space describes characteristics of table changes that consist of two dimensions, \ie, data objects and properties of data changes.
        $103$ characteristics are included in the space.
        The blue and orange square indicates data objects in the output and input table, respectively. A detailed description of each characteristic is provided in the supplemental material.}
	\label{fig:design_space_chart}
\end{figure*}

By analyzing the characteristics of table changes, we construct the design space consisting of two primary dimensions, \ie, data objects and the property of data changes.
Data objects include four different types, \ie, Tables, Columns, Rows, and Cells.
The first three data objects are recognized in various prior work in data wrangling\cite{kasica2020table,kai2022somnus} and table comparison\cite{niederer2017taco}. 
We introduce Cells additionally because it relates to a wide range of transformations including editing the text of a cell and filling empty cells with adjacent cells.
The second dimension is the property of data change. 
We identify five high-level properties, \ie, Number, Order, Relation, Value, and Type.
Intersections between data objects and properties mean a set of characteristics, which are described by inequalities.
For example, the intersection between Tables and Number includes five characteristics, \ie, the number of input and output tables goes from ``$0\rightarrow 1$'', ``$1\rightarrow 0$'', ``$1\rightarrow 1$'', ``$1\rightarrow \textmd{many}$'', and ``$\textmd{many} \rightarrow 1$'', as illustrated in Fig.~\ref{fig:design_space_chart}.
We notice that the mapping between data objects and properties is not complete, meaning that some properties are only applicable to specific data objects.
In the following paragraphs, we describe the five properties in detail.

\subsubsection{Number}
Numerous transformations can result in a change in the number of data objects including \textit{create tables}, \textit{delete columns}, etc. 
Number is applicable to tables, columns, and rows, as shown in Fig.~\ref{fig:design_space_chart}, 
For tables, the comparison between the number of inputs and outputs is critical.
For example, the \textit{create tables} transformation may increase the number of tables from $0$ to $1$. 
Meanwhile, the \textit{separate tables} may generate more tables based on one input table.
Therefore, we include five characteristics in this category, \ie, the number of input and output tables goes from ``$0\rightarrow 1$'', ``$1\rightarrow 0$'', ``$1\rightarrow 1$'', ``$1\rightarrow \textmd{many}$'', and ``$\textmd{many} \rightarrow 1$'', which corresponds to characteristics 1---5 in Fig.~\ref{fig:design_space_chart}.

For columns and rows, we summarize characteristics based on single- or multiple-table operations.
In terms of single-table transformations, we focus on whether the number of columns/rows in the output table is greater than (\ie, $>$, the \nth{8} characteristic in Fig.~\ref{fig:design_space_chart}), equal to ($=$), or smaller than ($<$) the number in the input table.
In terms of multiple-table transformations, changes in the number of columns/rows may vary. 
For example, considering the \textit{inner join} transformation, the number of rows in the output table is smaller than or equal to the minimum number of rows in the input tables.
On the contrary, the \textit{split table by groups} transformation may split a table row-wise.
And the number of rows in the input table equals the sum of rows in the output tables.
For detailed categorization, we further divide multiple-table transformations into two cases, \ie, $1\rightarrow \textmd{many}$ (one input table and many output tables) and $\textmd{many} \rightarrow 1$.
For the $1\rightarrow \textmd{many}$ case, we compute the minimum, maximum, and sum of the number of columns/rows in all output tables, and compare them to the number in the input table (\eg, characteristics 9---15). 
For the $\textmd{many} \rightarrow 1$ case, the same computation is performed for input tables, which is compared to the number of outputs (\eg, characteristics 16---22).
Each case includes seven characteristics, which are described in Fig.~\ref{fig:design_space_chart}.

\subsubsection{Order}
The order property describes the position change of columns and rows, which can be performed manually or be sorted according to criteria including disorder, ascending, and descending~\cite{kasica2020table}. 
When performed manually, the index of columns/rows is changed (\eg, characteristics 40 and 41).
When sorted based on criteria, the content of the output column/row may be ordered (\eg, characteristics 42---44).

\subsubsection{Relation}
Some data transformations involve relations between data objects (\eg, columns, rows, and cells). 
Taking the \lstinline|unite| function in \kai{tidyr} as an example, 
\lstinline|df2 = unite(df1,`Z', X, Y)| forms the column \lstinline|Z| in \lstinline|df2| by concatenating strings of columns \lstinline|X| and \lstinline|Y| in \lstinline|df1|.
Relations can be triggered by mathematical formulas and functions.
However, the relation space between data objects is too large to be exhausted.
By analyzing real-world wrangling scripts, we focus on four commonly-used set relations, \ie, subset  ($\subset$), superset ($\supset$), equal ($=$), and \kai{others (\ie, apart from these three relations, $\neq$)}.

For relations between columns (or rows), we summarize within- and between-table characteristics, respectively.
In terms of between-table characteristics, we focus on whether the content of columns in the input table is a subset ($\subset$), superset ($\supset$), identical ($=$), or \kai{others} ($\neq$) to that of the output table (\eg, characteristics 50---53). 
In terms of within-table characteristics, we mainly focus on identical ($=$) relations.
That is, we first identify the existence of identical columns within input and output tables, and compare them using boolean operators (\eg, characteristics 54---57). 
For relations between cells, the characteristics are described by whether the values are equal ($=$), substring (\eg, $\subset$ and $\supset$), or \kai{others} ($\neq$) to each other (\eg, characteristics 66---69).
\kai{Besides, we check whether cells in the same column or row have identical values in the input and output tables (\eg, characteristics 70---77).}


\subsubsection{Value}
Some transformations relate to special values, \eg, missing values and user-defined values.
For example, the \lstinline|dropna| function in Pandas detects and removes rows with missing values.
Meanwhile, the \lstinline|replace| function replaces cells with user-defined values.
The value property is applicable to all data objects. 
For tables, columns, and rows, we establish characteristics by identifying the existence of missing values within input and output data objects, and compare them using boolean operators (\eg, characteristics 78---81).
In addition, we detect whether the data objects include ($\supseteq$) a user-defined value or not ($\nsupseteq$).
As for cells, the characteristics are described by whether a cell is a missing value, and whether a cell has changed from one value to another.

\subsubsection{Type}
This property, which describes changes in data type, is applicable to columns.
We focus on three types of columns, \ie, nominal, quantitative, and temporal. 
The characteristics are described as, given a column in the input table, whether its data type is changed in the output table.

\subsection{Inferring Transformations With Characteristics}
\label{subsubsec:characteristic_library}
We summarize the characteristic space for the purpose of inferring the type of data transformation.
We begin with collecting a set of transformations.
Then we discuss the principles of type inference.

Kasica \etal\cite{kasica2020table} presented a design space encompassing $15$ high-level transformations, which provides a good start for our investigation. 
However, some transformations are too rough to distinguish detailed semantics. For example, constructing columns manually, mutating values from existing columns, and merging multiple columns, are all summarized by \textit{Create Columns} without distinction. 
\kai{On the other hand, Wrangler\cite{guo2011proactive} and Trifacta~\cite{Trifacta} provide vocabularies describing low-level operations, such as \textit{Extract}, \textit{Merge}, and \textit{Replace}.
To support finer-grained semantics, we manually annotate each line of code in our corpus (see Sec.~\ref{subsec:space_method}) by linking one high-level transformation and one low-level operation.}
Taking the above three transformations as examples, the new vocabularies of them are \textit{create\_columns\_create}, \textit{create\_columns\_mutate}, and \textit{create\_columns\_merge}, respectively.
This strategy is effective in generating a large number of low-level transformations.
In addition, we observe that some wrangling code does not change data tables, such as \lstinline|group|, \lstinline|ungroup|, and \lstinline|reset_index| functions in Pandas.
Though not changing data tables immediately, these functions usually act as prerequisites for the following transformations.
Hence, we include an \textit{identical\_operation} transformation to describe these functions.

To infer transformation type, we establish a mapping between transformations and characteristics.
For each transformation, we categorize characteristics into three groups, \ie, impossible, possible, and inevitable.
Impossible characteristics refer to those that should not appear when performing a transformation.
Inevitable characteristics, on the other hand, will definitely exist after a transformation.
The above two sets play major roles in characteristic-based inference (see Sec.~\ref{subsubsec:transformation_inference}).
Possible characteristics are those that may or may not appear.
For example, given a line of code, \lstinline|df=df[df.num>1]|, which removes rows not meeting the condition \lstinline|df.num>1|, the number of columns will not change (\eg, the characteristic 6 is inevitable), and the number of rows in the output table can be smaller or equal to that in the input table (\eg, characteristics 23 and 24 are possible).
However, changes in the number of columns (\eg, characteristics 7 and 8) are impossible.
As a result, the problem of inferring transformation type is turned into detecting a set of impossible, possible, and inevitable characteristics.


\section{The \name{} Pipeline}
This section introduces \name{}, an automatic pipeline for inferring semantics for data wrangling scripts.
Our pipeline takes data tables and a wrangling script as input, and generates a sequence of data transformations with parameters.
As illustrated in Fig.~\ref{fig:teaser}, \name{} consists of three key steps, \ie, Data Preprocessing (Fig.~\ref{fig:teaser}(b)), Type Inference (Fig.~\ref{fig:teaser}(c)), and Parameter Inference (Fig.~\ref{fig:teaser}(d)).

\subsection{Data Preprocessing}
\label{subsec:prepro}
Since table changes highly relate to the semantics of transformation, the goal of this step is to identify a set of characteristics of table changes.
To this end, we first execute the wrangling script on the given source table, and obtain intermediate input and output tables for each line of code, which are saved as CSV files for further analysis.
\kai{These input tables are critical for type inference as one line of wrangling code could have different semantics given different input tables~\cite{kai2022somnus}.}
\kai{Second, we detect table changes based on the design space for each wrangling code and output a set of characteristics.}
We notice that the detection is time-consuming since some characteristics require content-based comparison between two tables. 
For example, detecting the relation of two columns involves a pair-wise comparison of the content in each cell.
Therefore, we employ two pruning strategies to facilitate the detection.

\textit{Characteristics Pruning}: 
Most detection relies on the existence of input and output tables. 
Hence, a number of characteristics can be pruned if the prerequisite does not stand.
For example, if only output tables exist, the transformation is likely to be 
\textit{Create Tables}, and the detection of other characteristics, such as changes in order and relation, is not applicable.

\textit{Data Objects Pruning}: 
We observe that, given new data objects in the output table, characteristics are mainly caused by the new ones.
Hence, we prune the remaining objects for content-based comparison. 
For example, after detecting new columns in the output table, we focus on the detection of characteristics based on the new columns. 
Other columns in the output tables, however, will not be involved in the content-based comparison.

\subsection{Type Inference}
To infer the transformation type of a line of wrangling code, the Type Inference step includes two components, \ie, characteristic-based inference and model-based inference.
As the space of transformations could be large, the first component filters out impossible transformations and results in a smaller number of candidates.
The second component utilizes a Siamese convolutional neural network to assess the possibility of each candidate and returns a sorted list based on possibilities.

\begin{figure}[!ht]
    \centering
    \includegraphics[width=\linewidth]{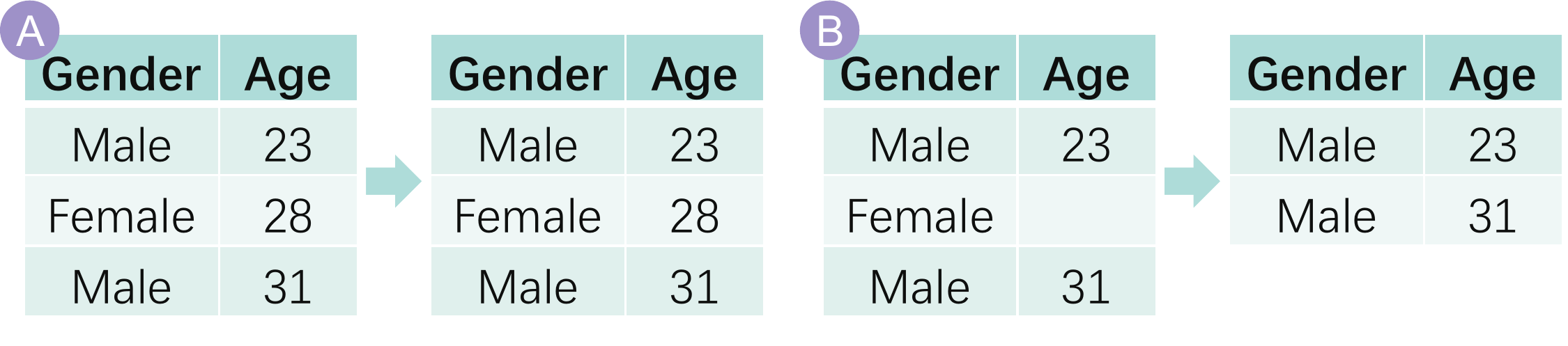}
    \vspace{-4mm}
    \caption{Examples of the two ambiguities in characteristic-based inference. (a) The output table is identical to the input table, which can be caused by \textit{Remove Duplicate Rows}, \textit{Remove Rows with Missing Values}, \textit{Sort Age in Ascending Order}, etc. (b) The second row has been deleted, which can be caused by performing \textit{Filter Rows Where Gender is Male}, \textit{Remove Rows with Missing Values}, etc.}
    \label{fig:ambiguities}
    \vspace{-0.26cm}  
\end{figure}

\subsubsection{Characteristic-Based Inference}
\label{subsubsec:transformation_inference}
\kai{Given a collection of characteristics,}
we filter out impossible transformations based on the mapping described in Sec.~\ref{subsubsec:characteristic_library}.
We establish two main strategies.
First, we identify whether characteristics fall into the \textit{impossible} set of each transformation.
If true, the transformation is not included in the candidate list.
Second, we filter out transformations whose \textit{inevitable} set is not a subset of the set of characteristics.
In this component, the type of transformation can hardly be uniquely identified because of two kinds of ambiguities.
First, the output table may be identical to the input one.
This may be because the conditions of transformations do not meet.
An example is shown in Fig.~\ref{fig:ambiguities}(a), the input table may be unchanged after \textit{Remove Duplicate Rows} because no duplicates exist in the input. 
However, the same result may be caused by \textit{Remove Rows with Missing Values} or \textit{Sort Age in Ascending Order}, etc.
Second, more than two transformations lead to the same changes between tables.
As Fig.~\ref{fig:ambiguities}(b) shows, changes between two tables may be caused by \textit{Filter Rows Where Gender is Male}, \textit{Remove Rows with Missing Values}, etc.
Hence, this component results in a set of transformation candidates.

\subsubsection{Model-Based Inference}

\kai{
To resolve the ambiguities in characteristic-based inference, the analysis of scripting code seems a feasible solution.
However, due to the variety of programming languages, functions, and parameters in data wrangling, crafting rules to parse code is time-consuming and limited in generalizability and scalability.
Besides table changes, we observe that the semantics of a transformation is usually implied in the code.
For example, some functions, \eg, \lstinline|drop_duplicates| in Pandas, \lstinline|distinct| in dplyr, and \lstinline|unique| in R, perform \textit{Removing Duplicate Rows}, which can be inferred from the function name to some extent.
It is also true for some parameters, such as \lstinline|ascending=False|, \lstinline|remove=TRUE|, etc.
Therefore, we employ a machine learning model to alleviate ambiguities and compute the possibilities for each transformation candidate.}
One key step in machine learning is feature engineering, which identifies critical information and generates vectors for it.
Here we incorporate the characteristics, function names, and parameters as key information for feature engineering.

To reveal the semantics of function names and parameters, we vectorize them using text embedding approaches.
The embedding, however, is not straightforward.
To be specific, some wrangling code does not include a function name.
Fig.~\ref{fig:Somnus_examples}(c) presents an example where a new column in the output table is created by an addition statement.
In this case, we set \lstinline|None| as its function name.
In addition, function names and parameters usually contain out-of-vocabulary (OOV) words.
For example, some functions are named \lstinline|drop_duplicates|, \lstinline|inner_join|, and \lstinline|fillna|, and some parameters are called \lstinline|by_group|, \lstinline|to_replace|, and \lstinline|skipna|.
To tackle this problem, we adopt FastText~\cite{bojanowski2016enriching} for text embedding because it can handle the OOV words by considering subword information.
Finally, we obtain two $1 \times 300$ vectors capturing the semantics of function names and parameters, respectively.
Similarly, we present each characteristic using natural language, and use FastText to vectorize each characteristic.
After obtaining all feature vectors of characteristics, we average all vectors to form one with dimension $1 \times 300$.
Finally, we linearly combine these vectors to obtain a semantic vector with dimension $3 \times 300$.

To calculate the likelihood of each transformation, we employ a Siamese Convolutional Neural Network (CNN)~\cite{Aditya2019Siamese} which can address two challenges. 
First, the problem of inferring transformation type is relatively new, and training samples are hard to find in the existing corpus.
Second, the data transformation space could be large.
However, only a few of them are used commonly in practice (details are described in Sec.~\ref{subsec:dataset}).
That means training samples from the real-world corpus are likely to be imbalanced.
CNN is a kind of flexible classification architecture that is suitable for few shot problems and imbalanced data \cite{koch2015siamese}.
We acknowledge that \name{} does not rely on one model architecture, and the CNN model can be replaced by others. 
In this paper, we use CNN to verify the idea of \name{}.

The Siamese network model comprises two identical subnetworks, and each utilizes five convolutional layers followed by a flatten layer.
The input of each subnetwork is a semantic vector encompassing three types of information, \ie, the characteristics of changes, the function name, and the function parameters. 
In this model, we refer to Hadsell \etal\cite{hadsell2006dimensionality} to define and optimize the contrastive loss.
After a pair of semantic vectors is fed into the model, the similarity of the two vectors is computed based on Euclidean distance.
The output of the model is $1$ or $0$, which means whether the two inputs belong to the same transformation type.
\kai{After training the model on the labeled dataset (see Sec.~\ref{subsec:dataset}), we apply it to obtain the likelihood of each transformation in the candidate set.}


\subsection{Parameter Inference}
\label{subsec:para}
Parameters define how a transformation performs.
\kai{Taking the set of transformation candidates with likelihood as input, this step aims to infer parameters for transformations.}
Borrowing the idea of slot filling in task-based dialog systems \cite{shi2020sequence}, we first manually define a set of parameters, \ie, slots, that a transformation should include.
Then, we detect these parameters, \ie, filling slots, based on the wrangling code, characteristics of table changes, and the content of input/output tables.
For example, given a piece of code, \lstinline|df2 = unite(df1, `Z', X, Y)|,
\name{} infers that the transformation \textit{merge columns} has the maximum likelihood.
Then, we detect parameters of \textit{merge columns}, which are pre-defined as one input table, multiple columns in the input, one output table, the merged column in the output, and a separator.
The input and output tables can be obtained from the Data Preprocessing step.
To fill other slots, we first extract column names from the code, \ie, \lstinline|Z|, \lstinline|X|, and \lstinline|Y|.
Then, we obtain relations between the three columns using characteristics of table differences, and fill the two slots, \ie, multiple columns in the input and the merged column in the output.
Since the separator is not available in the code, we extract it from the table content.
Note that \name{} only supports the inference of a limited number of separators, and can be easily extended.

However, the slot filling process \kai{may fail}.
If so, \name{} will filter out the transformation and turn to the one with the second maximum likelihood and detect whether parameters meet the transformation, and so forth.
\name{} stops detection until parameters match the transformation type.
Following the above example, if relations between columns cannot be detected, the transformation type, \eg, \textit{merge}, does not stand.
Then, \name{} will turn to the next transformation and continue the slot filling process.
This step can also filter out impossible transformations to some extent, which is beneficial to the overall accuracy of \name{}.

\section{Experiments}
\label{experiment}
\kai{We design experiments to evaluate \name{} from two aspects.
First, since characteristic-based inference is independent of programming languages, we assess whether the pipeline trained in one programming language can be generalized to the other. Second, we evaluate how each part of \name{} contributes to the overall performance.}

\begin{table}[!ht]
    \centering
    \caption{ 
      \kai{Statistical analysis of the dataset on five metrics, including the total lines of wrangling code (\#Instances), the average lines of wrangling code in scripts (Avg \#Inst), the number of distinct data transformation types (\#Trans), 
      the number of distinct functions (\#Func), and the average number of implementations for each data transformation (Avg \#Imps).}
    }
    \resizebox{\linewidth}{!}{\begin{tabular}{l|ccccc}
        \toprule
        &     \textbf{\#Instances} & \textbf{Avg \#Inst} & \textbf{\#Trans} & \textbf{\#Func} & \textbf{Avg \#Imps} \\ \midrule
        Python  & 606 & 14.43 & 29 & 52 & 3.86 \\   
        R       & 315 & 9.84 & 28 & 47 & 2.14 \\ \midrule  
        Sum & 921 &  & 30 & 99 &   \\
        \bottomrule        
    \end{tabular}}
    \label{tab:dataset_analysis}
\end{table}

\subsection{Dataset}
\label{subsec:dataset}
Based on the corpus described in Sec.~\ref{subsec:space_method}, we manually annotate the transformation type for each line of code.
To ensure the accuracy, the labeling process is based on the documentation of each function describing its usage.
After that, we verify the labeling results using the characteristics between the input and output tables.
Finally, we have labeled and verified $74$ script files including $921$ lines of code.
We distinguish scripts written in Python and R, and report the statistical results of our dataset on five metrics (see Table~\ref{tab:dataset_analysis}).
\kai{
Here we define the number of implementations (\#Imps) as how many methods (including different functions or the non-functional equation) can be utilized to perform a data transformation.
For example, there are two functions (\ie, \lstinline|pivot_longer| and \lstinline|gather|) in tidyr that can be used to fulfil \textit{transform\_tables\_fold}, so its \#Imps is $2$.}

From Table~\ref{tab:dataset_analysis}, we have some interesting findings.
Though we have tried to balance the number of scripts ($42$ for Python and $32$ for R), the number of instances between the two languages varies ($606$ versus $315$). It is because Python scripts are usually longer than those of R ($14.43$ Avg \#Inst for Python versus $9.84$ for R).
The number of transformations and functions that appeared in the corpus is similar.
\name{} can support $99$ functions in total, which is significantly more than previous work including Datamations\cite{pu2021datamations} and Sonmus\cite{kai2022somnus}.
We argue that our pipeline is scalable in terms of the number of functions.
However, we notice that the Avg \#Imps in Python ($3.86$) is greater than that in R ($2.14$), meaning that Python is more flexible in implementing data transformations.
\kai{The dataset and its statistical analysis are provided in the supplemental material.}


\subsection{Metrics}
Following the previous work~\cite{shankar2020evaluating}, we adopt Top-N accuracy metrics to evaluate the performance of \name{}. 
These metrics are usually used for multi-class classification models.
Top-1 accuracy is a conventional classification metric that measures the proportion of samples where the highest probability prediction matches the target label.
However, due to a large number of categories (\ie, $30$ distinct transformations) and a few training/testing samples (\ie, $921$ instances), Top-1 accuracy may limit our understanding of \name{}.
Hence, we employ Top-3 accuracy, which measures the possibility that the top three predictions include the correct answer.
\kai{We also report the training times in minutes for each experiment to get a rough idea about how computationally-demanding the training process is.}

\subsection{Methods and Apparatus}

To evaluate how each component contributes to the performance, we prepare two techniques, \ie, the entire pipeline and the Siamese Convolutional Neural Network (CNN).
Similar to \name{}, the CNN technique includes three steps.
The difference between them is that the second step of the CNN technique does not include a characteristic-based inference component.

We split the dataset using four strategies.
\kai{Firstly, to evaluate how each technique performs across Python and R, 
we use the scripts in Python as the train set and R as the test set, and vice versa.}
To ensure that the space of data transformations is consistent between the train and test, we select  common transformations ($27$ types in total) in both sets.
Secondly, we split the Python instances by random selection in each transformation type at a ratio of $80:20$, which are used as the train and test sets, respectively.
Based on the Python test set, we form the third strategy where we include R scripts into the train set.
Finally, we combine all Python and R instances and randomly split them at a ratio of $80:20$.
The experiments are performed on a Windows 10 desktop with a 3.20GHz Intel Core i7-8700 CPU and 32 GB RAM.

\subsection{Results and Analysis}

\begin{table}[ht]  
  
    \centering  
  \caption{The training time, Top-1, and Top-3 performances of \name{} and its CNN-based component in different experiment settings.}
  \vspace{-0.2cm}
  \resizebox{\linewidth}{!}{
    \begin{threeparttable}  
    \label{tab:experiment_results}  
      \begin{tabular}{cccccccc}  
      \toprule  
      \multirow{2}*{} & \multicolumn{2}{c}{Setting} &  
      \multicolumn{2}{c}{CNN} & \multicolumn{2}{c}{\name{}} & Time\cr  
      \cmidrule(lr){2-3} \cmidrule(lr){4-5} \cmidrule(lr){6-7}  
      & Train & Test & Top-1 & Top-3 & Top-1 & Top-3 & (minutes)\cr  
      \midrule  
      \nth{1} & Python & R & 19.1 & 30.3 & 53.8 & 76.1 & 19.5 \cr  
      \nth{2} & R & Python & 23.2 & 32.3 & 44.5 & 82.6 & 15.1 \cr  
      \nth{3} & Python & Python & 62.1 & 83.3 & 80.3 & 94.7 & 18.3\cr  
      \nth{4} & All & Python & 72.0 & 90.9 & 90.2 & 98.5 & 21.9 \cr  
      \nth{5} & All & All & 78.0 & 92.2 & 92.2 & 99.0 & 19.6 \cr  
      \bottomrule  
      \end{tabular}  
      \end{threeparttable}  
    }
    \vspace{-0.4cm}
  \end{table}

\textit{Contribution of each component:} 
The Type Inference step of \name{} consists of two components, \ie, characteristic-based inference and CNN-based inference. 
By incorporating the characteristic-based inference component, the pipeline results in better accuracy compared to the CNN-based model on both Top-1 and Top-3 accuracies.
\kai{Specifically, based on the Top-1 accuracy, the accuracy increases by $21.32\%$ on average ($\sigma =7.06$) for the five data splits.}
It is also notable that CNN-based inference achieves acceptable performance for the last three data splits, indicating that both components have a great contribution to the pipeline.

\textit{Generalizability to different languages:} 
To understand the generalizability, we focus on the \kai{first two experiment settings}.
\kai{Generalization across Python and R is a hard task. 
\name{} achieves $53.8\%$ and $44.5\%$ Top-1 accuracy for Python-R and R-Python, respectively,
while the accuracy of CNN-based model is lower, merely $19.1\%$ and $23.2\%$, respectively.}
We speculate the low accuracy is due to the small number of train samples which are not enough for the model to capture features.

\textit{Rationality of semantic vector:} 
By comparing the results of the \nth{3} and \nth{4} data splits, we infer that the embedding of function names and parameters is significant in type inference.
Specifically, by incorporating R scripts into the train set, the semantics in R has joined for inference,
which improves the Top-1 accuracy of both CNN-based model and \name{} by $9.9\%$.
This finding inspires us for improving the pipeline design in future research, where we plan to include synonyms for each transformation in model training.

\textit{Analysis of Confusion Matrix:} 
Referring to the normalized confusion matrix (provided in the supplemental material), we observe that 
\textit{create\_rows\_summarize} (\ie, creating rows by summarizing other rows with aggregate operations such as mean, sum, and max) tends to be predicted as \textit{create\_rows\_create}, which means rows are created manually. 
Another case is that some \textit{transform\_columns\_extract} (\ie, transforming columns by extracting values from one column) are classified as \textit{transform\_columns\_mutate}.
We infer two reasons for the two cases. 
First, characteristics in our design space cannot describe complex relations such as summarization and regular expressions.
Second, the instances of these transformations are few, making it difficult for \name{} to learn their features.


\subsection{Potential Improvements}
\label{subsec:improve}
Although \name{} has achieved $92.2\%$ of the Top-1 accuracy when combining R and Python scripts in the train and test sets, the analysis of failure cases informs us of potential improvements in our pipeline.

First, the semantics of parameters should be investigated by advanced approaches.
Currently, we use FastText to convert all parameters into one feature vector, which may overlook key semantics.
For example, by analyzing a failure case, \lstinline|df4 = filter(df3,!is.na(ST))|, it should be classified as \textit{delete\_rows\_dropna}, while \name{} incorrectly predicted it as \textit{delete\_rows\_filter}.
This case is ``difficult'' because the two transformations are semantically similar, \ie, deleting rows with missing value is a kind of deleting rows using filtering conditions.
By analyzing the code, we note that the parameter \lstinline|!is.na(ST)| is a key factor to distinguish it from \textit{delete\_rows\_filter}.
The failure case indicates that the Type Inference step of \name{} has not fully revealed the semantics of parameters.
\kai{We plan to investigate natural language processing techniques such as tokenization~\cite{kagkelidis2021lumina, luo2021natural, wang2019vispubcompas, domingues2022legalvis} to address this issue.}


Second, characteristics detection can be further augmented.
The characteristic-based inference component is powerful in improving the performance of the pipeline.
However, given a pair of input and output tables, characteristics detected in the current prototype are sometimes ambiguous for type inference (as described in Fig.~\ref{fig:ambiguities}).
We observe that the detection of characteristics highly relies on the input data tables.
\kai{Thus, to resolve ambiguity, we plan to employ well-crafted data tables to verify each transformation candidate. For example, if \textit{delete\_rows\_deduplicate} is one of the candidates, a well-designed input table with duplicate rows would be constructed to check whether \textit{delete\_rows\_deduplicate} stands for the line of code.}



\section{Example Applications}
\label{sec:app}
In this section, we demonstrate the \kai{flexibility} of \name{} through three applications in different domains.
First, we show how \name{} supports semantic annotation for scripts in Jupyter Notebooks.
Second, we integrate \name{} into an existing program visualization system to augment its ability to parse wrangling scripts.
Third, we demonstrate an idea of improving an existing visualization system with additional information obtained from \name{}.

\subsection{Code Annotation in Jupyter Notebook}
\label{subsec:annotations}

\begin{figure}[ht]
	\centering
	\includegraphics[width=\linewidth]{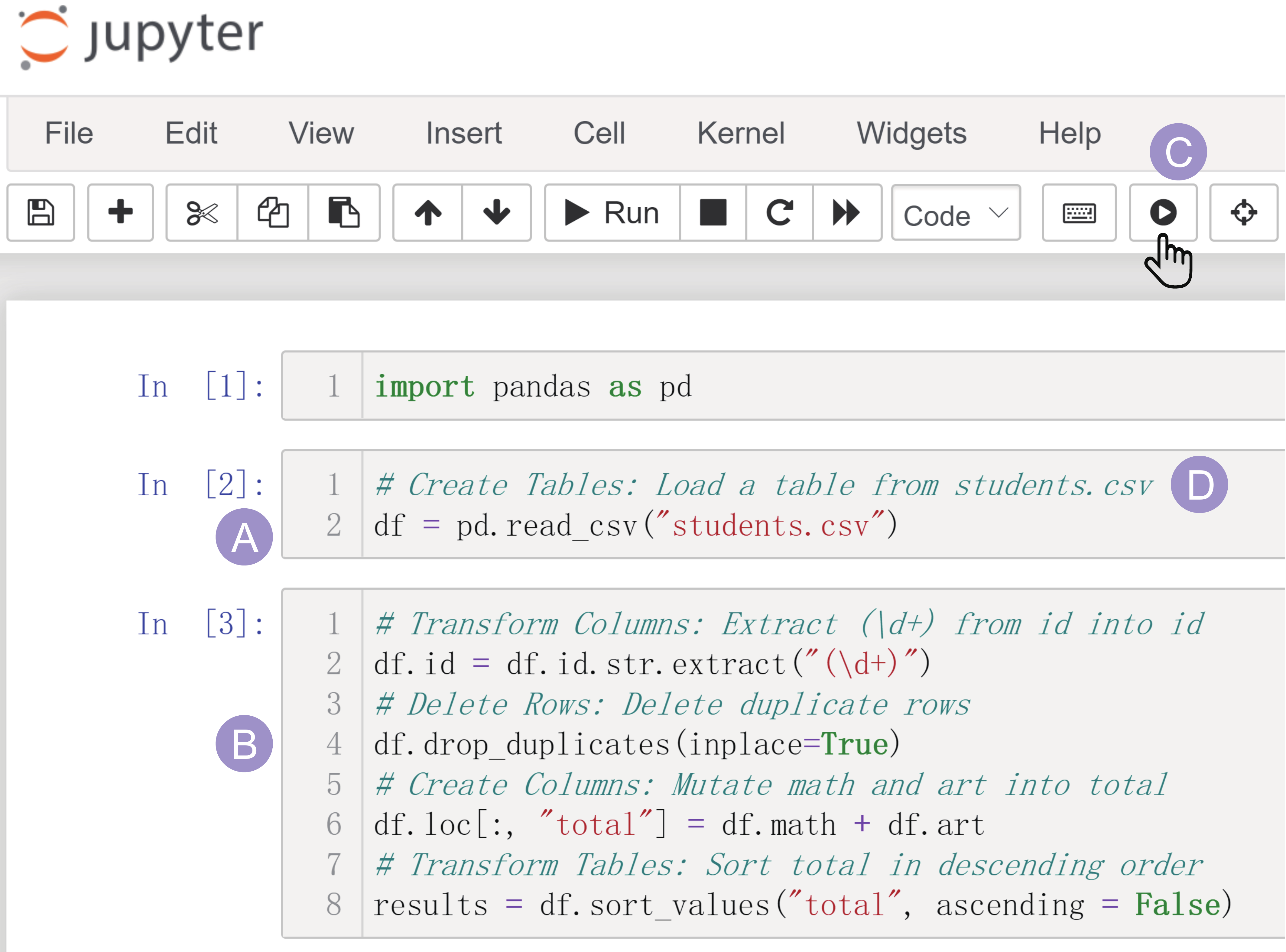}
	\caption{An instance of generating annotations for data wrangling code in Jupyter Notebook. After wrangling code has been written \kai{in two code cells (a, b)}, the annotations \kai{(d)} describing transformation semantics are appended to the original code by clicking the extension button \kai{(c)}.}
	\label{fig:annotations_JN}
\end{figure}


Computational notebooks leverage an interactive literate programming~\cite{knuth1984literate} paradigm that combines code, natural language text, and execution results of code cells in a single document.
Nowadays, increasing data workers adopt them for exploratory data analysis and sharing of computational narratives~\cite{perez2015project}.
However, writing comments for each line of code could be burdensome.
To assist the processing of documenting for data workers, we implement \name{} as a Jupyter Notebook extension to generate semantic annotations for wrangling code snippets.
Specifically, to obtain intermediate input and output tables, we utilize the Variable Inspector~\cite{variableinspector} to collect all variables with their types, sizes, and values.
After \name{} infers the transformation type and its parameters of a line of wrangling code, the extension employs a template-based approach to generate a natural language sentence describing the semantics.
And the sentence will be appended above to the original code as a comment.
As presented in Fig.~\ref{fig:annotations_JN}(a) and (b), there are five lines of wrangling code in the script, which are distributed in two different code cells.
After clicking the extension button (Fig.~\ref{fig:annotations_JN}(c)), comments will be automatically inserted above each line of code (Fig.~\ref{fig:annotations_JN}(d)).
As code annotation is a hot topic in the domain of software engineering \cite{hu2018deep, li2020deepcommenter}, \name{} shows potential for code annotation in data wrangling.

\subsection{\somnus{} Re-visited}

\begin{figure*}[!htbp]
	\centering
	\includegraphics[width=\linewidth]{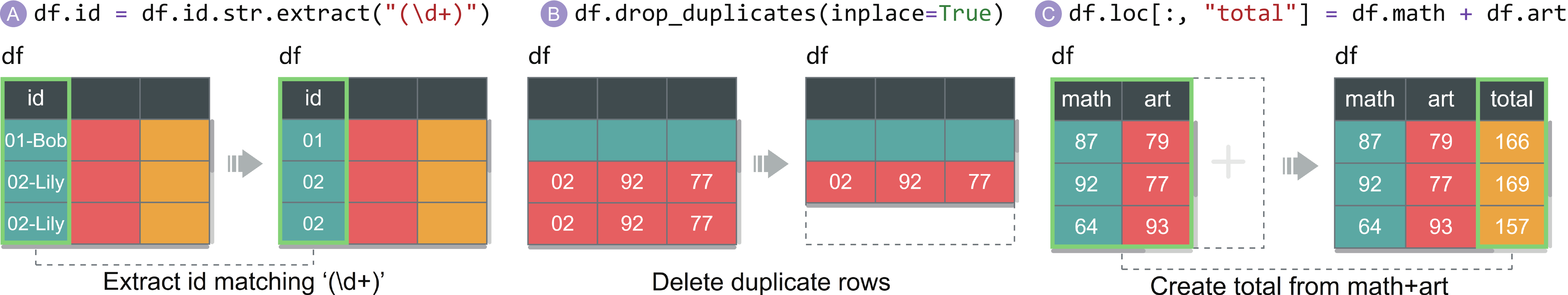}
	\caption{Three examples that \somnus{} can support with the help of \name{}. Before integrating \name{}, \somnus{} does not support (a) the \lstinline|extract| function, (b) non-assignment statements, and (c) non-functional assignments.}
	\label{fig:Somnus_examples}
	\vspace{-0.1cm}
\end{figure*}

\name{} can also be used to expand the capacity of existing visualization systems that reveal the semantics of transformations.
For example, \somnus{}~\cite{kai2022somnus} is a program visualization system that generates a provenance graph to visualize a script of data wrangling.
However, it relies on hand-crafted rules to parse code and infer transformation types.
Hence, the backend of the system is limited in scalability.
To augment \somnus{}, \name{} replaces its Program Adaptor module that parses code and infers the type of transformation.
As a result, \somnus{} supports a larger number of functions, and more types of statements (\eg, non-functional assignments and non-assignment statements).
Fig.~\ref{fig:Somnus_examples} shows three examples that \somnus{} can support with the help of \name{}.
In Fig.~\ref{fig:Somnus_examples}(a), the \lstinline|extract| function was not supported in \somnus{}, and establishing rules for new functions is time-consuming. 
Fig.~\ref{fig:Somnus_examples}(b) shows that the non-assignment statement was not supported because the input and output table is not explicitly presented.
Fig.~\ref{fig:Somnus_examples}(c) depicts a non-functional assignment that can be hardly mapped to manually-crafted rules.
\kai{Our corpus (see Sec.~\ref{subsec:space_method}) shows that 507 out of 921 ($55.05\%$) instances match the above three scenarios.}
By incorporating \name{}, \somnus{} extends its ability to deal with a wider range of functions and statements. 

\subsection{TACO Re-visited}

TACO \cite{niederer2017taco} is a visual analytics system that facilitates table comparison.
It presented four types of changes between two tables including 1) \textit{structural changes} where rows or columns are added or removed, 2) \textit{content changes} where the values of cells are modified, 3) \textit{reorder changes} where rows or columns are shifted to other positions, and 4) \textit{merge/split changes} where multiple rows or columns are combined into a single one or vice versa. 
Though similar, our design space is the superset.
That is, our space includes characteristics that are not supported in TACO, such as inclusion relationships between columns/rows (\eg, characteristics 50---53), ordering states (\eg, characteristics 42---44), and data types (\eg, characteristics 102 and 103).
Therefore, \name{} can assist TACO to reveal more characteristics of table changes.

\begin{figure}[!h]
    \centering
    \includegraphics[width=\linewidth]{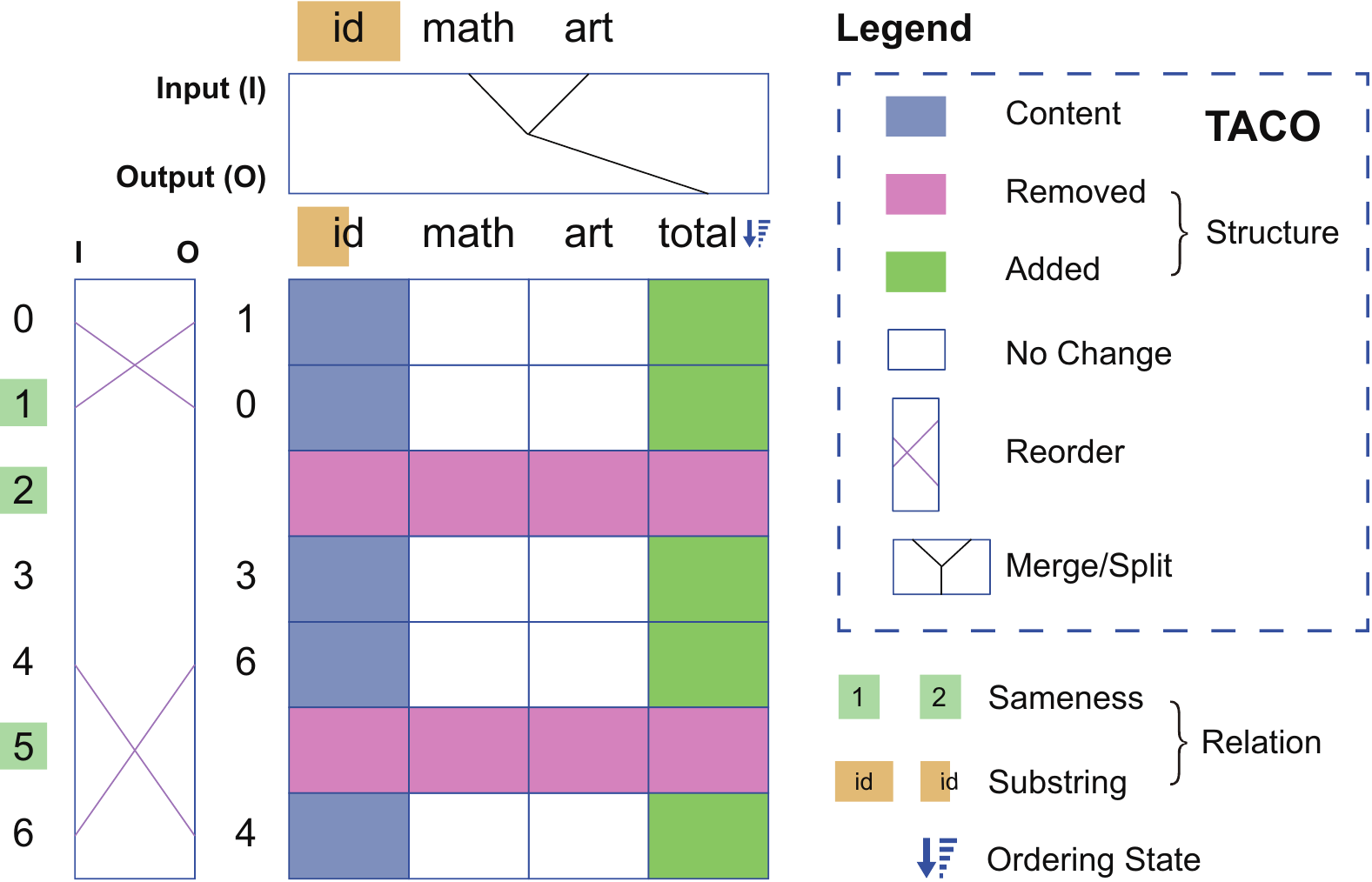}
    \caption{A diff heatmap visualization of table differences generated by a modified version of TACO based on the characteristics of table changes detected from \name{}. The visual encodings enclosed by the dashed box in the legend are from TACO.}
    \label{fig:semfer_taco}
	\vspace{-0.3cm}
\end{figure}

We express this idea using a simplified but extended version of TACO.
In this version, we use additional visual channels to represent two relationships, \ie, green markers on the left side encode duplicate rows, and brown markers on top of the heatmap to display the superset relations between two columns.
Fig.~\ref{fig:semfer_taco} shows an example.
Based on the original visual encodings, we observe two red rows, which indicate that they are deleted.
However, it is not clear why they are removed.
With the help of \name{}, we can infer the reason with relation information between columns and rows.
For example, we can observe that there are three green markers on the left side, meaning that there are identical rows in the input table.
Combining with the two red rows, we may infer that a \textit{deduplicate} transformation is performed to remove the two rows.
Besides, two brown markers appear on the top side, which indicates the \lstinline|id| between input and output tables have superset relation. 
A descending symbol is placed close to \lstinline|total|, which indicates that the \lstinline|total| column is sorted in descending order.
By extending TACO, we can observe detailed relations between two tables.
An exciting research direction is exploring visual encodings or novel designs to reveal richer characteristics of table changes effectively and intuitively.

\section{Discussion}
In this section, we first discuss the limitations of our pipeline and the experiments. Then we present potential usage for future exploration.

\subsection{Limitations}

\kai{\textbf{Evaluation.} The evaluation of \name{} is limited in three aspects.
To begin with, the quantitative experiments show that our pipeline could infer the transformation type with high Top-N accuracy. However, we did not assess the quality of the inferred parameters, which are critical in understanding semantics as well.
In addition, we have not tried models other than the Siamese Convolutional Neural Network in Model-Based Inference.
Lastly, the design space of table changes has not been explicitly evaluated in terms of descriptive, evaluative, and generative power.
In future research, we plan to design comprehensive experiments to assess the quality of \name{} and the design space, and investigate how \name{} performs with different multi-class classification models.}

\kai{\textbf{Generalizability.}}
Generalization across \kai{Python and R} is difficult, and the current prototype only achieves 
\kai{$53.8\%$ accuracy for Python/training and R/test, and $44.5\%$ for R/training and Python/test.}
When performing the same transformation, function names and parameters in different languages may vary while characteristics between tables are the same.
We anticipate that the key factor to improve the generalizability is to align the semantics of function names and parameters among different languages.
In addition, a large number of training and test samples are also critical for models to parse the semantics.

\kai{\textbf{Scalability.}}
Compared to prior work \cite{pu2021datamations,kai2022somnus}, \name{} scales well in terms of the number of functions and parameters.
However, our pipeline does not scale well in the number of transformation types.
The mapping between transformation types and characteristics (described in Sec.~\ref{subsubsec:characteristic_library}) and rules for parameter inference (see Sec.~\ref{subsec:para}) are crafted manually for each transformation.
Our pipeline currently supports $30$ types of transformations that are derived from the collected dataset.
However, the transformation space has \kai{not} yet been fully explored.
Crafting rules for the entire space could be tedious and error-prone.
In future iterations, we plan to investigate automatic approaches to build rules for the pipeline.

\kai{\textbf{Capability.} The limitations of \name{} in capability are three-fold.}
First, \name{} infers one transformation with its parameters for a line of wrangling code.
It is unable to reveal the semantics of a line of code containing multiple transformations.
\kai{Second, it is not applicable to scripts with some imperative programming constructs, such as conditional statements and loops, which are commonly used in data wrangling.
Third, it does not support function chaining due to the difficulty of obtaining intermediate data tables.
Temporarily, this issue could be addressed by automatically converting the chained functions into individual lines.}
The three capability issues are critical to production usage, and we plan to address them in our future work.


\subsection{Opportunities for New Usage}
\label{subsec:opportunity}
With the ability to infer  the semantics of wrangling scripts, \name{} can introduce opportunities for practical applications in various domains.
\kai{Apart from the three example applications described in Sec.~\ref{sec:app}, we anticipate that our work can be integrated into productivity spreadsheet software and interactive data wrangling systems, including Microsoft Excel~\cite{MicrosoftExcel} and OpenRefine~\cite{OpenRefine}, to automatically record the history of table operation. 
This is helpful for documenting and sharing the transformation process and the table changes, and the history can be presented to augment the ``undo'' operation and used for auditing transformations~\cite{kandel2011research}.}
In such a scenario, wrangling code is not explicitly generated.
To adapt to these systems, feature engineering for function names and parameters should be transformed to the understanding of operation history in the interface.

Another possible usage of \name{} is to support a source-to-source compiler that converts code from one programming language to another~\cite{roziere2020unsupervised}.
Numerous source-to-source compilers have been proposed~\cite{roziere2020unsupervised, tangiblesoftware, transpiler}, which usually require large amounts of samples for training and testing.
These techniques, however, can hardly support code translation in the domain of data wrangling for two reasons.
First, samples for data wrangling are usually not well-documented, and manual annotation is laborious and time-consuming. 
Hence, learning-based approaches may result in poor performance without enough training samples.
Second, subtle differences in the wrangling code may lead to different transformations.
Learning the mapping among numerous functions and parameters across different languages is infeasible.
\name{} narrows the gulf by aligning wrangling code with semantics.
To be specific, we first infer the semantics for individual wrangling functions and their parameters using \name{}.
Then, we establish bi-directional mappings between functions/parameters and semantics for different programming languages.
Finally, the semantics act as bonds connecting functions in two languages.

\section{Conclusion and Future Work}
In this paper, we have presented \name{}, a three-step pipeline that infers semantics for wrangling scripts.
\name{} takes a piece of wrangling script and data tables as input, and outputs the semantics of each line of code consisting of the type of transformation and its parameters.
Based on the observation that differences between input and output tables highly relate to the transformation type.
We summarize a design space presenting characteristics of table changes, which further guides the design of \name{}.
Experiments suggest that \name{} performs well in type inference.
Further, three applications in different domains indicate that \name{} has good \kai{flexibility}. 

In the future, we plan to explore four promising research directions.
First, we plan to improve the performance of \name{} by refining the Data Preprocessing step and Type Inference module.  
Second, we want to enhance its capability to meet the need for production usage.
Third, we would like to incorporate \name{} in a broad range of applications, including spreadsheet software, data wrangling tool, and source-to-source compiler in the domain of data wrangling.
Fourth, we plan to further evaluate \name{} with \kai{comprehensive experiments}.

\acknowledgments{
  The work was supported by NSFC (62072400, 62002331) and the Collaborative Innovation Center of Artificial Intelligence by MOE and Zhejiang Provincial Government (ZJU). This work was also partially funded by the Zhejiang Lab (2021KE0AC02).}

\bibliographystyle{abbrv-doi}

\bibliography{template}
\end{document}